\documentclass[12pt,oneside,a4paper,english]{article}
\usepackage[T1]{fontenc}
\usepackage[latin2]{inputenc}
\usepackage[margin=2.25cm,headheight=26pt,includeheadfoot]{geometry}
\usepackage[english]{babel}
\usepackage{listings}
\usepackage{color}
\usepackage{titlesec}
\usepackage{titling}
\usepackage{changepage}
\usepackage{amsmath}
\usepackage{hyperref}
\usepackage{enumitem}
\usepackage{graphicx}
\usepackage{fancyhdr}
\usepackage{lastpage}
\usepackage{caption}
\usepackage{tocloft}
\usepackage{setspace}
\usepackage{multirow}
\usepackage{titling}
\usepackage{float}
\usepackage{comment}
\usepackage{booktabs}
\usepackage{indentfirst}
\usepackage{lscape}
\usepackage{booktabs,caption}
\usepackage[flushleft]{threeparttable}
\usepackage[english]{nomencl}
\usepackage{xcolor}

\title{BIOREME Study Group Report} 
\makeatletter\let\Title\@title\makeatother

\newlist{steps}{enumerate}{1}
\setlist[steps, 1]{leftmargin=1.5cm,label = Step \arabic*:}

\newcommand{\fref}[1]{Fig. \ref{#1}}

\begin{document}
\pagenumbering{roman} 

\pdfoutput=1
\begin{titlepage}
\begin{center}
\vspace{1.0cm}
\includegraphics[width=0.4\textwidth]{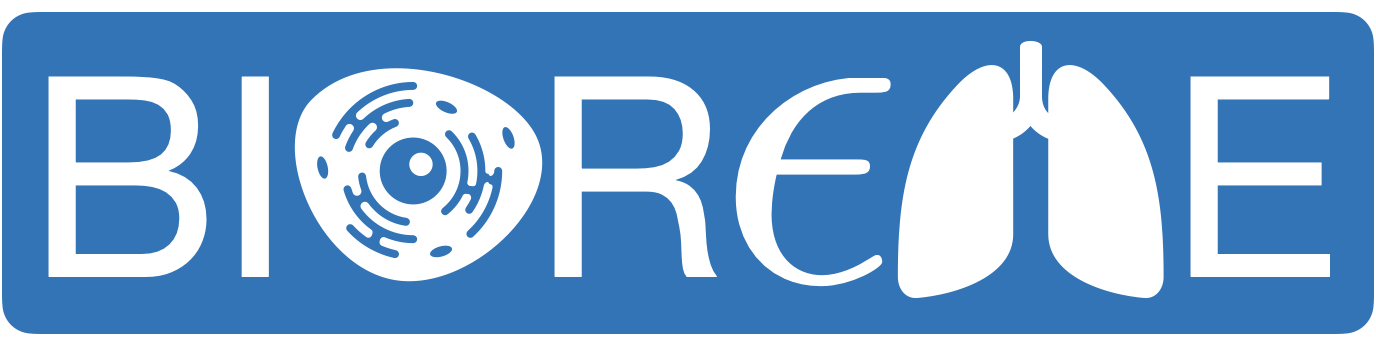}~\\[1cm]

{\large\textbf{Developing the next generation of lung function measurement:
A Study Group in Mathematical Modelling}}
\vspace{.5cm}

\vspace{2.0cm}

\hrule
\vspace{.5cm}
{\huge\textbf{Correction and standardisation of lung oscillometry techniques using parameter inference: A study group report}} 
\vspace{.5cm}

\hrule
\vspace{1.5cm}

\textsc{\textbf{Authors}}\\
\vspace{.5cm}
\centering

Bindi S. Brook -- University of Nottingham\\
Graeham R. Douglas -- Arete Medical Technologies\\
Oliver E. Jensen -- University of Manchester\\
Sonal Mistry -- University of Warwick\\
Sujit Kumar Nath -- University of Manchester\\
Matthew J. Russell -- University of Nottingham\\
Sina Saffaran -- University of Warwick\\
James Shemilt -- University of Manchester\\
Liam Weaver -- University of Warwick\\
Carl A. Whitfield* -- University of Manchester\\

\vspace{1cm}
*Corresponding author: \href{mailto:carl.whitfield@manchester.ac.uk}{carl.whitfield@manchester.ac.uk}\\
General enquiries: \href{mailto:contact@bioreme.net}{contact@bioreme.net}

\vspace{2cm}

\centering Date of publication: \today 
\end{center}
\end{titlepage}

\newpage
\doublespacing
\renewcommand{\baselinestretch}{1}\normalsize
\tableofcontents
\renewcommand{\baselinestretch}{1}\normalsize
\thispagestyle{fancy} 

\newpage
\pagenumbering{arabic} 
\fancyfoot[C]{Page \thepage\ of \pageref{EndOfText}}

\section{Introduction} 
\label{sec:intro}

This report relates to a study group hosted by the EPSRC funded network, Integrating data-driven BIOphysical models into REspiratory MEdicine \href{https://www.bioreme.net/about}{(BIOREME)}, and supported by \href{https://www.sheffield.ac.uk/insigneo}{The Insigneo Institute}  and \href{https://iuk.ktn-uk.org/industrial-maths/}{The Knowledge Transfer Network}. The BIOREME network hosts events, including this study group, to bring together multi-disciplinary researchers, clinicians, companies and charities to catalyse research in the applications of mathematical modelling for respiratory medicine. The goal of this study group was to provide an interface between companies, clinicians, and mathematicians to develop mathematical tools to the problems presented. The study group was held at The University of Sheffield on the 17 - 20 April 2023 and was attended by 24 researchers from 13 different institutions. Below details the technical report of one of the challenges and the methods developed by the team of researchers who worked on this challenge.

\subsection{The Challenge}
\label{sec:challenge}

\textbf{Background:} Arete Medical Technologies are developing a respiratory medical device, ``The Respicorder'' (\href{http://www.aretemedtech.com}{http://www.aretemedtech.com}) that combines several measurements of lung function into one portable device. One of these measurements uses impulse oscillometry (IOS), whereby a small puff of air is quickly oscillated into and out of the mouth during normal breathing. The resulting pressure and flow rate changes can be used to the impedance of the airways, which in turn can provide proxy measurements for (patho)physiological changes in the small airways.

\textbf{The problem:} Disentangling the signal so that airway mechanics can be measured accurately (and device properties/environmental effects can be accounted for) remains an open challenge that has the potential to significantly improve the device and its translation to clinic.

\textbf{Data available:} Participants had temporary access to IOS measurements and data from the Forced Oscillation Technique (FOT), collected by Arete for the purposes of this workshop.

\subsection{Proposed Solution}
\label{sec:solution}

We propose that one way to standardise FOT/IOS interpretation is to provide biophysical model-based analysis tools that can infer the RLC parameters\footnote{I.e. the resistance, inductance, and capacitance in an electrical analogy, see section \ref{sec:circuit-models} for further explanation.} of different components of the system (i.e. device, extra-thoracic and lung). These components could be discerned using data collected from different calibration tests (e.g. device-only data, valsalva and k-manouevres) in addition to the standard FOT/IOS patient test data. 

We also propose that these mechanistic physical models then provide a unique opportunity to couple data from multiple datasets even with relatively small patient numbers. Multi-modal data integration may provide one way to get even more insight and robustness out of multiple lung function probes, such as those that the Arete device can perform. 

\section{Methods} 
\label{sec:methods}

The team broadly split into two workstreams, one working directly with the data provided (sections \ref{sec:device_data}), and the other developing models of the lung contribution to the oscillometry signal (sections \ref{sec:airway-models}). 

\subsection{Identifying device effects from the data}
\label{sec:device_data}

It was found that attempts to infer parameters of the device were significantly impacted by a near-constant offset (or `drift') in the measured flow, so this had to be corrected first.

\subsubsection{Correcting the data for flow offset}
\label{sec:correction}

\begin{figure}[h!]
\centering
\includegraphics[angle=0,scale=0.5]{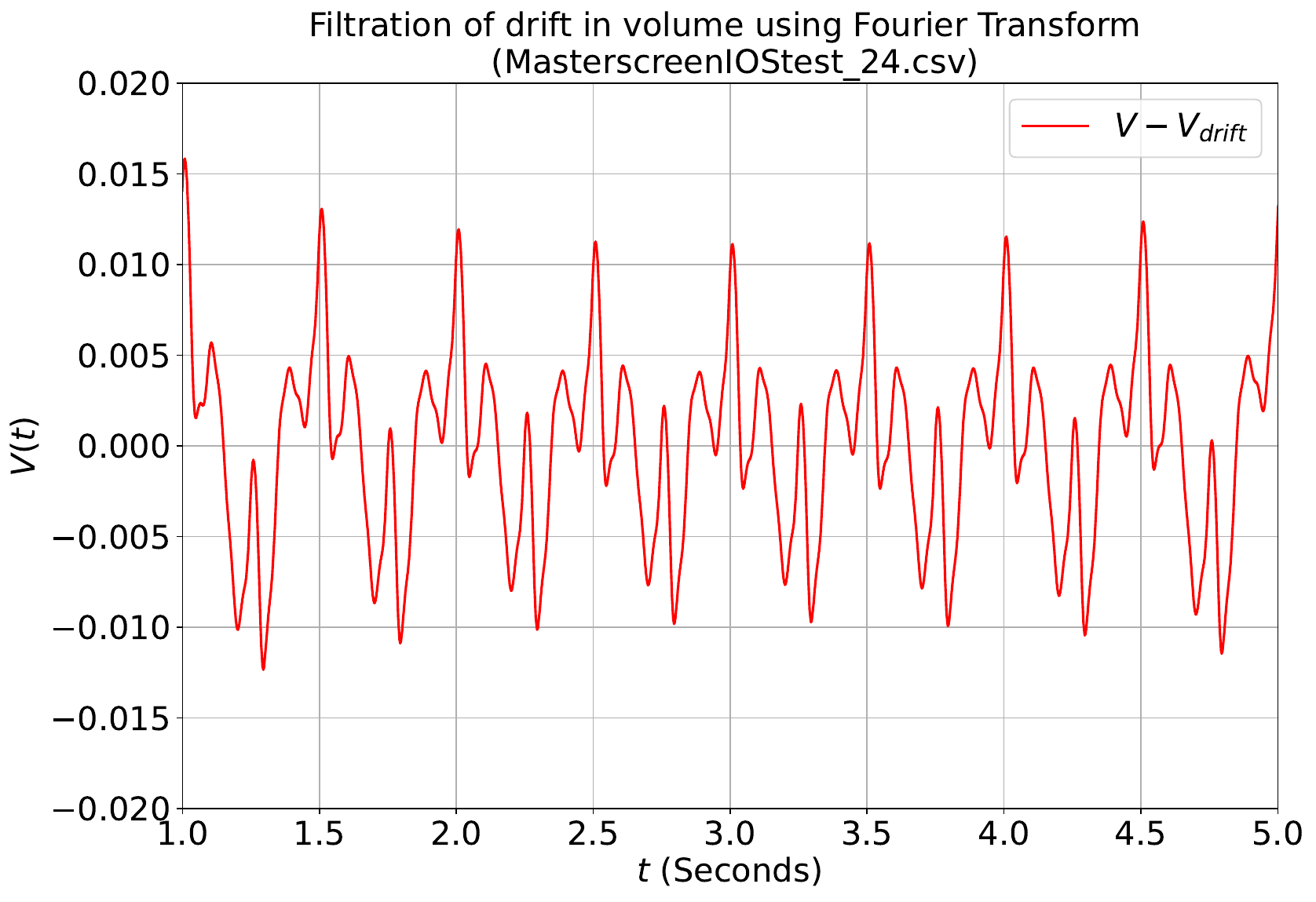}
\caption{A section (in steady state) of the drift-filtered volume data.}
\label{figVolFiltSS}
\end{figure}
\begin{figure}[h!]
\centering
\includegraphics[angle=0,scale=0.5]{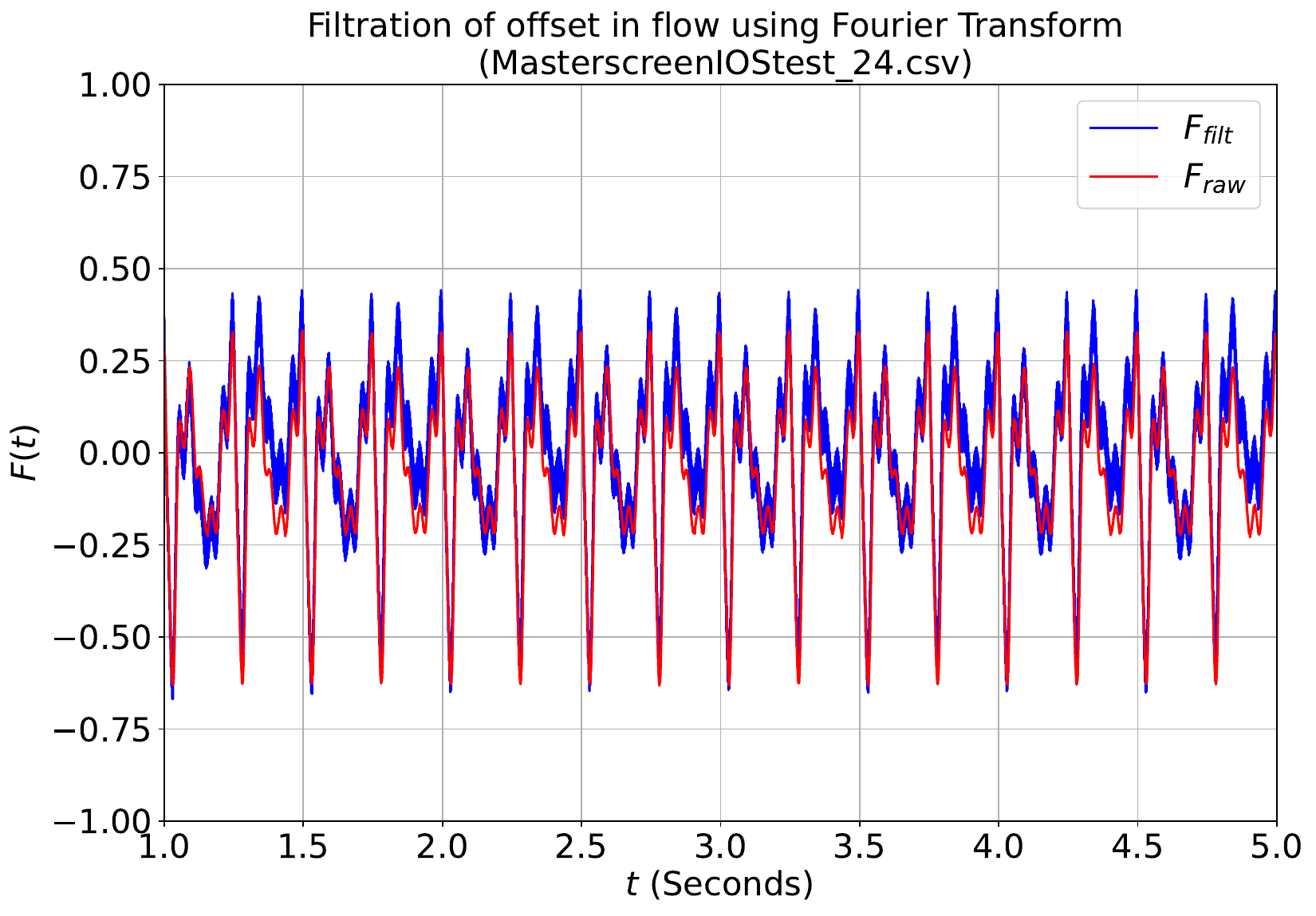}
\caption{Filtration of offset from the flow data. Red is the raw flow data contaminated
by some unknown offset. Blue is the flow data obtained by taking derivative of the drift-free
volume data.}
\label{figFlowFiltSS}
\end{figure}

Other than subtracting the average drift from the volume data by fitting straight lines,
which we already have found to be very effective, we also propose another method
for filtering the drift with the help of Fourier transform. \\
%
\begin{figure}[h!]
\centering
\includegraphics[angle=0,scale=0.5]{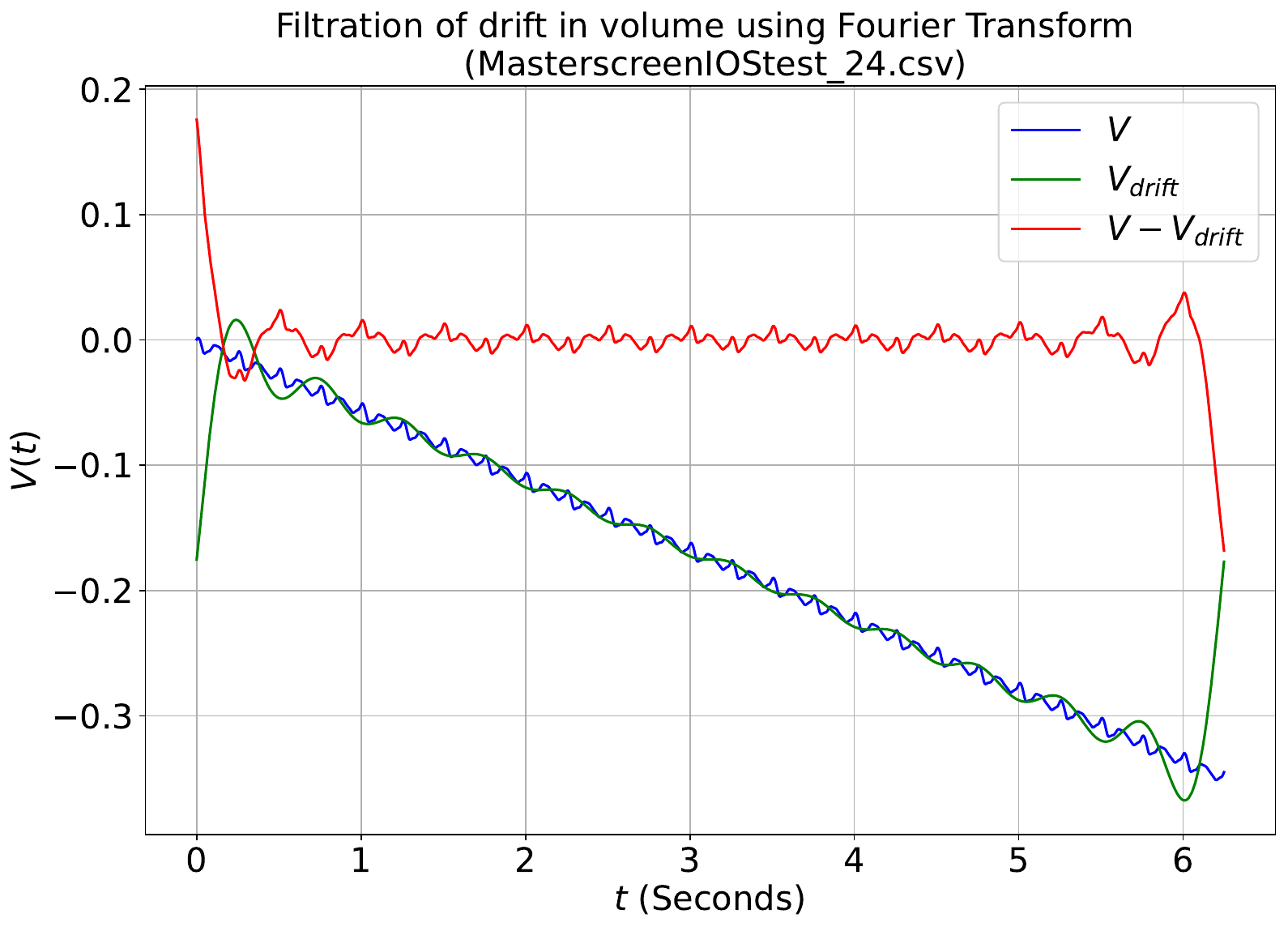}
\caption{Filtration of drift from the volume data. The blue line is the raw volume data.
Green is the drift in the data contributed by the low frequency Fourier modes. Red curve
is obtained by subtracting the green from the blue curve representing the volume data
without drift.}
\label{figVolFiltFourier}
\end{figure}
%
\fref{figVolFiltFourier} shows an example of such filtration. In this method we
first calculate the low frequency modes of the volume data which give
rise to the drift we see in the data. Then we subtract those low frequency modes from the original
volume data to make it drift-free. Assuming the data is in the steady state, we can now choose
any subsection of the data which is away from the boundaries of the time interval on which the
Fourier analysis is performed (\fref{figVolFiltSS}). Finally, taking the derivative of the
drift-free volume data we can recover the flow data without having any offset as shown
in \fref{figFlowFiltSS}.
%
%
%
%
\clearpage

\subsubsection{Inferring parameters of circuit models of the device}
\label{sec:circuit-models}
The real-life processes can be reproduced and better understood by means of modelling and engineering approaches. It also can be engaged in testing and evaluating hypotheses. The human pulmonary system comprises all organs and structures within the body that play a significant role in the process of respiration. Respiration begins at the nose or mouth, where oxygenated air is brought in. The trachea is the starting point to transfer the gas. It is the largest of all the airways and at its distal end branches into two bronchi leading to the left and right lung. Each bronchus progressively branches into shorter, narrower airways called bronchioles. Bronchioles subsequently branch into elastic cavities called alveoli. The lungs consist of approximately 300 million alveoli, where different gasses are transferred into and out of the bloodstream. The factors that determine the gas flow to and from the alveolar units are the airway resistances and the pressure gradient between the mouth and the lungs.

The system of airways can be explained by means of electrical analogy. In these terms, any airway would be represented as a wire and a number of impedances to represent the characteristics of the airway (e.g. narrow vs wide, compliant vs rigid), the flow of air would be the electrical current and the pressures are shown as the voltage in the circuit. Along the same lines, a mechanical ventilator can be treated as a voltage or current source depending on the mode of ventilation we are using.

\begin{figure}[h!]
\centering
\includegraphics[angle=0,scale=0.5]{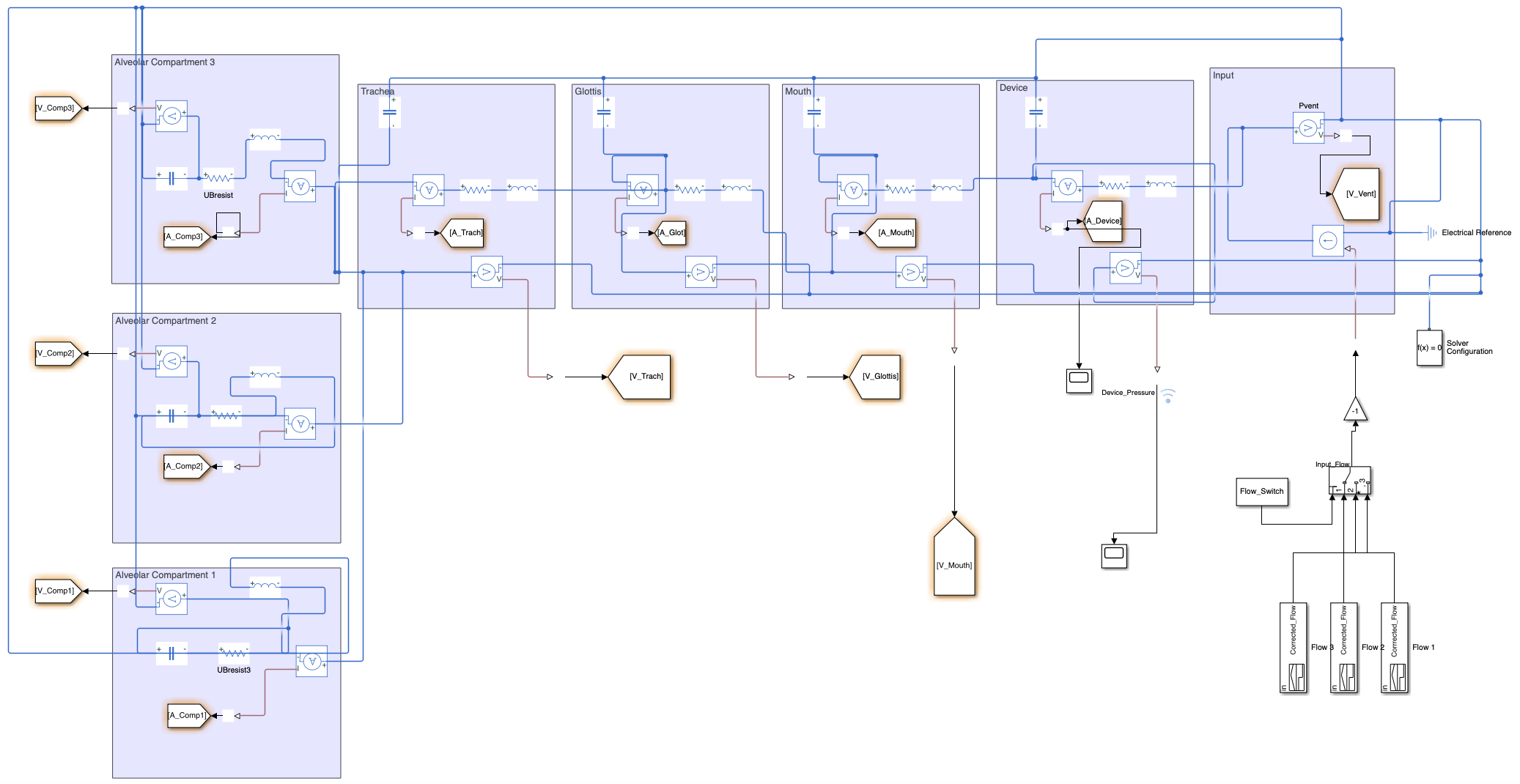}
\caption{Figure shows an outline of the designed model on Simulink. The model includes 3 alveolar compartments as well as a single compartment for each trachea, glottis, mouth and device}
\label{figCircuitModelOutline}
\end{figure}

The respiratory system can be described by means of multi-compartmental models. A model is effective when studying different features such as compliance, resistance, ventilation-perfusion distribution and delivery of gas across the whole lung. It also offers the ability to consider inhomogeneity in the lung. For the purpose of this study, we developed an electrical analogy to build a simple model of the respiratory airways as well as a few alveoli to be able to parameterise them and estimate the network impedance using the pressure and flow data from impulse oscillometry. \fref{figCircuitModelOutline} shows an outline of the designed model on Simulink. The model includes 3 alveolar compartments as well as a single compartment for each trachea, glottis, mouth and device. Different flow profiles from the data are fed into the model as the input. The Genetic Algorithm (GA) is employed as the optimisation tool to find a set of model parameters (i.e. the impedance) that best matches the pressure outputs of the model to the corresponding pressure data.

\subsection{Mathematical models of the respiratory system}
\label{sec:airway-models}

We begin by outlining the physics of oscillatory flow in a single airway and the resulting system of equations for the whole airway network in sections \ref{sec:osc_flow} and \ref{sec:tree_traversal}. Then in appendix \ref{sec:network-models} we show how, by considering the balance of inertial and elastic forces, the eigenmodes of the resulting linear system can reveal how different frequencies excite different regions of the airway tree. 

We also discuss circuit models of airway impedance in the literature and use these to compute total impedance in some image-based airway networks (section \ref{sec:real_networks}), before proposing a simplified model that can be incorporated into the circuit models introduced in section \ref{sec:inverse_models}.

\subsubsection{Oscillatory flow in a single airway}
\label{sec:osc_flow}

Oscillatory flow in airways is often modelled using the classical solution for laminar unsteady flow in an infinitely long tube \cite{foy_computational_2017, bhatawadekar_modelling_2015}. Assuming axisymmetry and unidirectionality, the Navier-Stokes equation reduces to
\begin{align}
\label{NSeq} \rho \frac{\partial u}{\partial t} = \mu \frac{\partial}{\partial r}\left(r\frac{\partial u}{\partial r}\right) - \frac{\partial p}{\partial x}
\end{align}
where $u(r,t)$ is the flow velocity along the tube axis $x$, $p(x,t)$ is the pressure assumed constant over a given cross-section of the tube, $r$ is the radial coordinate, and $t$ is time. Lack of $x$ dependence implies the pressure gradient is constant $\partial p/\partial x = \Delta P(t) / L$. As we are interested in oscillatory flow we consider solutions of the form $\Delta P/L =  (\Delta P_0/L)e^{i \omega t}$ and $u = u_0 e^{i \omega t}$. Substituting into equation \eqref{NSeq} we get
\begin{align}
\mu r \frac{{\rm d}^2 u_0}{{\rm d} r^2} + \mu \frac{{\rm d} u_0}{{\rm d} r} - \rho i \omega u_0 - (\Delta P_0/L) = 0.
\end{align}
This is a form of Bessel equation, assuming a no-slip boundary at $r=R$ the solution is
\begin{align}
\label{plug_velocity} u_0 = \frac{\Delta P_0}{\rho L i \omega}\left[1 - \frac{J_0 (\alpha \sqrt{-i} r/R)}{J_0 (\alpha \sqrt{-i})} \right],
\end{align}
where $J_0(x)$ is the Bessel J function and $\alpha = R \sqrt{\rho \omega / \mu}$ is the dimensionless Womersley number. Thus, the volumetric flux through the tube is given by  $Q = Q_0 e^{i\omega t}$ where $Q_0$ is determined by integrating equation \eqref{plug_velocity} over the tube cross-section \cite{womersley_method_1955-1,thurston_periodic_1952}
\begin{align}
    \label{Zadef} Q_0 = \frac{\pi R^4 \Delta P_0}{i \mu L \alpha^2}\left(1 - \frac{2J_1(\alpha\sqrt{-i})}{\alpha \sqrt{-i} J_0(\alpha \sqrt{-i})}\right) \equiv \frac{1}{Z_a} \Delta P_0.
\end{align}
The complex impedance $Z_a(R,L,\omega)$ is the impedance value we use for an airway of length $L$ and radius $R$. Note that for small $\alpha$ (i.e. small inertial contribution) $Z_a$ can be approximated as
\begin{align}
    Z_a \approx \frac{8 L \mu}{\pi R^4}\left[\left(1 + \frac{\alpha^4}{1152}\right) + i\frac{\alpha^2}{6} \right] + O(\alpha^6).
\end{align}
Thus, for small $\alpha$ (and recall that $\alpha \propto \omega^{1/2}$) the real part of the impedance is given by the Poiseuille resistance plus a modification proportional to $\omega^2$. The complex part (reactance) is proportional to $\omega$ and so this term gives the approximate `inertance' of the airway. These approximations could be used for the resistance and inertance terms referenced in appendix \ref{sec:network-models}. Note there are significant assumptions involved in this expression for the impedance. First, it assumes the the flow profile is a steady oscillating flow, which ignores the effects of boundary and initial conditions. Also, it assumes that airway deformation is negligible. 

\subsubsection{Tree traversal to calculate total airway network impedance}
\label{sec:tree_traversal}

If we make use of the circuit model analogy then the total airway impedance can be modelled by combining the individual airway impedances using Kirchoff's laws. Note the circuit analogy implies that the Wormersley flow is established in each branch and boundary layers/interactions between airways are insignificant. Given the significant inertial component this is a questionable assumption, and one that ought to be validated (e.g. against computational fluid dynamics simulations in branching geometries).

Even with this assumption, it can be a challenging task to calculate the total impedance at any given airway in the network as one needs to calculate the cumulative contribution from all the connected airways downstream in the network. Owing to the fact that each airway divides into exactly two branches, we propose a
binary tree model to uniquely define the serial number and location of each airway in the tree. In this model, we represent the airways as the nodes of the tree and assign them binary numbers systematically as follows.
%
\begin{figure}[h!]
\centering
\includegraphics[angle=-1.9,scale=0.5]{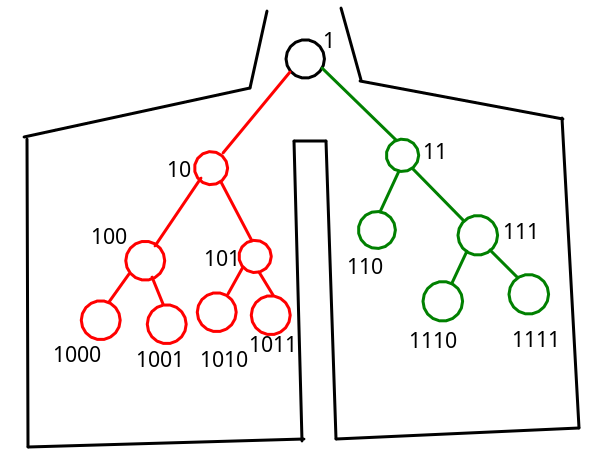}
\caption{Schematic diagram of the binary tree model of lung airway network.}
\label{figbinaryLung}
\end{figure}
%
The initial airway, trachea, is given the number $1$. The left primary
bronchus and right primary bronchus are assigned with binary numbers $10$ and $11$ respectively.
Every time an airway branches, the left daughter branch is assigned with a binary number
obtained by concatenating a $0$ on the right of the mother branch's string. The binary
number for the right daughter branch is obtained in the same way but by concatenating $1$.
To give an example, if the mother branch is assigned with a binary number $101$, then its left
daughter branch is assigned the number $1010$ and the right daughter branch is assigned the
number $1011$. The binary tree model of the lungs' airway network has some unique features as
the node numbers uniquely define their position in the tree. \\ \\
1. The second binary digit (from the left) of the string denotes which lung it is in. If the
digit is $1$, the node is in the right lung, e.g. $110$. Similarly, if the second binary digit
(from the left) is $0$, the node is in the left lung, e.g. $101$. \\
2. The number of digits in a node number denotes the generation of the airway. For example,
the node $1001$ is clearly a fourth generation node, considering the trachea in the first
generation. \\ \\
We can now calculate the impedance at any given airway by using the impedance formula
 %
 \begin{eqnarray}
 \label{eqnImpFormula}
Z_M = z + \frac{1}{\frac{1}{Z_L}+\frac{1}{Z_R}},
 \end{eqnarray}
%
where $Z_M$ is the total impedance of the mother airway, $z$ is the intrinsic impedance of each  individual airway, $Z_L$ and $Z_R$ are the total impedances of the left and 
right daughter airways respectively.

The algorithm of impedance calculation will be as follows. If we want to calculate
the impedance of an airway having its assigned binary string $[...]$, we first
list all the ancestors of this airway in the network by concatenating $0$ or $1$
sequentially to the right of the string. We do this until we hit the final generation.
Then we calculate the impedances of the airways by climbing up the tree generation
wise. $(n-1)^{th}$ generation impedances are evaluated using the impedances of
$n^{th}$ generation as
 %
 \begin{eqnarray}
 \label{eqnImpTree}
z_{[...]} = z + \frac{1}{\frac{1}{z_{[...]0}}+\frac{1}{z_{[...]1}}},
 \end{eqnarray}
%

Algorithms to automatically traverse trees in different ways are common in computer science. An example of how to execute the recursive computation in equation \eqref{eqnImpFormula} in Mathematica is to use the function \verb|TreeTraversalOrder| and specify \verb|"BottomUp"| ordering to guarantee that the cumulative daughter airway impedances are calculated before their respective parent's.

\subsubsection{Application to realistic airway networks}
\label{sec:real_networks}

Using equation \eqref{Zadef} for $Z_a$, in combination with the tree traversal method outlined in section \ref{sec:tree_traversal}, we calculate the total impedance of several model airway networks. The networks are the same as those used in \cite{whitfield2020} (and are available here \cite{whitfield_dataset_2020}). The image based networks are derived from CT scans of young people with Cystic Fibrosis but normal spirometry (original imaging study \cite{marshall_detection_2017}) and with little or no airway abnormality visible on CT. The vast majority of the airways in these models are generated by the space-filling branching algorithm outlined in \cite{whitfield2020} (a combination of those presented in \cite{tawhai_generation_2000,bordas_2015_developmet}), and thus, at baseline, are essentially models of healthy lungs. 

In order to incorporate the airway compliance, we follow the models in \cite{foy_computational_2017,bhatawadekar_modelling_2015} and assume that only the acini are compliant, and all have equal compliance. Therefore, each terminal airway in the model has a modified impedance of 
\begin{align}
    Z_{t} = Z_a - \frac{i}{\omega C_t},
\end{align}
where $C_t = C/N_t$ is the compliance of a single acinus, $C$ is the overall compliance of all the acini combined, and $N_t$ is the number of acini. 

We then study the effects of perturbations to these networks in the form of constrictions to the airways, in a similar manner to the approach in \cite{foy_computational_2017}. However, in this case the goal is to generate artificial data for simpler inverse compartmental models and to investigate which features of the physiology could be resolved by FOT or IOS measurements. We apply constrictions to the network by choosing airways within a certain range of Horsfield generations ($g_h$) at random and reducing their radius by 75\%. We define the generation ranges as "Proximal" ($g_h > x$), "Central" ($y < g_h \leq x$) and "Distal" ($g_h < y$), within which we constrict either 10\%, 25\% or 50\% of the airways. 

\subsubsection{Inverse models of lung impedance}
\label{sec:inverse_models}

A number of RLC-type circuit models of the lung have been proposed in the literature (see \cite{diong_augmented_2009} for a summary) which can extract the compliance and resistance of several lung components based on the frequency dependence of the measured impedance in FOT or IOS. However, none of these models consider the role of heterogeneity in lung properties that is common is obstructive lung disease, particularly asthma \cite{lui_role_2017}. Furthermore, parameters in these models are not always identifiable, so more robust parameter estimation and uncertainty quantification methods are required to help interpretation of results.

Here, we demonstrate a simple model to show how heterogeneity in respiratory resistance can alter the frequency dependence of the impedance, and some preliminary tests on parameter identifiability in this model. We propose a simple circuit model to represent airway heterogeneity where the airway component consists of $N$ resistors in parallel, with conductances drawn from a lognormal distribution: 
\begin{align}
\begin{split}
    R_i = R_0 / c_i \quad {\rm where} \quad c_i &\sim \mathrm{Lognormal}(-\sigma^2/2,\sigma).
\end{split}
\end{align}
This assumption is based on studies in ventilation heterogeneity where ventilation rates (and thus, roughly speaking, airway conductances) are seen to be well approximated by a lognormal model \cite{whitfield_model-based_2022,mountain_potential_2018}. Each resistor is connected in series to a capacitor with capacitance $1/(N K_0)$ (representing the alveoli with total elastance $K_0$) and grounded (representing the fixed pleural pressure). At the non grounded end, all of the airways are connected to single inductor with inductance $L_0$ which is connected in series to the source (the mouth). The frequency response then is determined by the parameters $L_0$, $K_0$, $R_0$ and $\sigma$, where $\sigma$ governs the heterogeneity in airway conductance. We test this model for parameter identifiability by generating measurements of impedance $Z$ and using Markov Chain Monte Carlo to estimate the parameters from this same generated data. The parameter priors are chosen as
\begin{align}
\pi(K_0) = U(1,200), \quad \pi(R_0) = U(0.5,50), \quad \pi(L_0) = U(0.001,0.5), \pi(\sigma) = U(0,2),
\end{align}
where the units are in terms of cmH2O, cmH2Os, and cmH2Os$^2$ for $K_0$, $R_0$ and $L_0$ respectively. Note that these parameters are unlikely to be sufficient to cover all of the identifiable lung mechanics features from this data, the goal here is just to test the feasibility of such a model.

This circuit model was then used to generate example data for a range of different (approximately physiological) parameter values for frequencies $\omega \in \{ \}$. The model was then fitted back to the data to find out which parameters were identifiable using a Markov Chain Monte Carlo algorithm.  The likelihood functions we use are
\begin{align}
L_R = {\rm Normal}(R_k,s_r^2), \quad {\rm and} \quad L_X = {\rm Normal}(X_k,s_x^2),
\end{align}
i.e. we assume that the error in resistance and reactance measurements are normally distributed with variances $s_r$ and $s_x$ respectively, where these parameters are also to be fitted by the MCMC algorithm. For the purposes of the parameter identifiability tests we assume priors of $s_r, s_x \sim U(0.001,0.01)$, so as to restrict to the case of very small measurement error. In future studies identifying parameters from real data, these priors will likely need to be much broader.

\section{Results} 
\label{sec:results}

\subsection{Preliminary results of fitting the circuit model to data}
\label{sec:device_data_results}
The first simulation which was run looked to characterise the device by considering its RLC Equivalent Circuit (EC), using the signal obtained from the device alone. The GA which was used to parameterise the EC failed to reach the termination criteria of minimal change to the cost function due to time constraints within the project, but it had begun to converge with the results indicating that the device is mainly resistive, with very little weighting given to the inductive and capacitive properties. The best result set obtained during the project characterised the device as having an equivalent resistance of 0.75 $\Omega$, an equivalent inductance of 0.03 H and an equivalent capacitance of 0.01 F. The tolerance of each of the above is 0.01 due to limitations with the Simulink Simscape block set used to perform the EC analysis.
Given more time, the model should converge on an optimal parameter set for the RLC values to characterise the device. The use of a GA was possibly unnecessary, with a less computationally expensive algorithm not suffering from a loss of performance due to the small size of the search space. It is suggested that the above results are taken as a benchmark and that a localised grid search takes place around these values to determine the most optimal parameter set for the device EC, especially as global optimisation algorithms such as a GA often fail to find the exact global minima but act to find the region within which it is contained. The next stage of this work would be to consider the other signals including patient and device, and using the known device characteristics, to eliminate the device effects effects from the signal, allowing a separation between the effects of the device on the signal and the signal resulting from the patient.

\subsection{Impedance in realistic lung airway networks}
\label{sec:impedance_real}

\begin{figure}
\centering
\includegraphics[width=\textwidth]{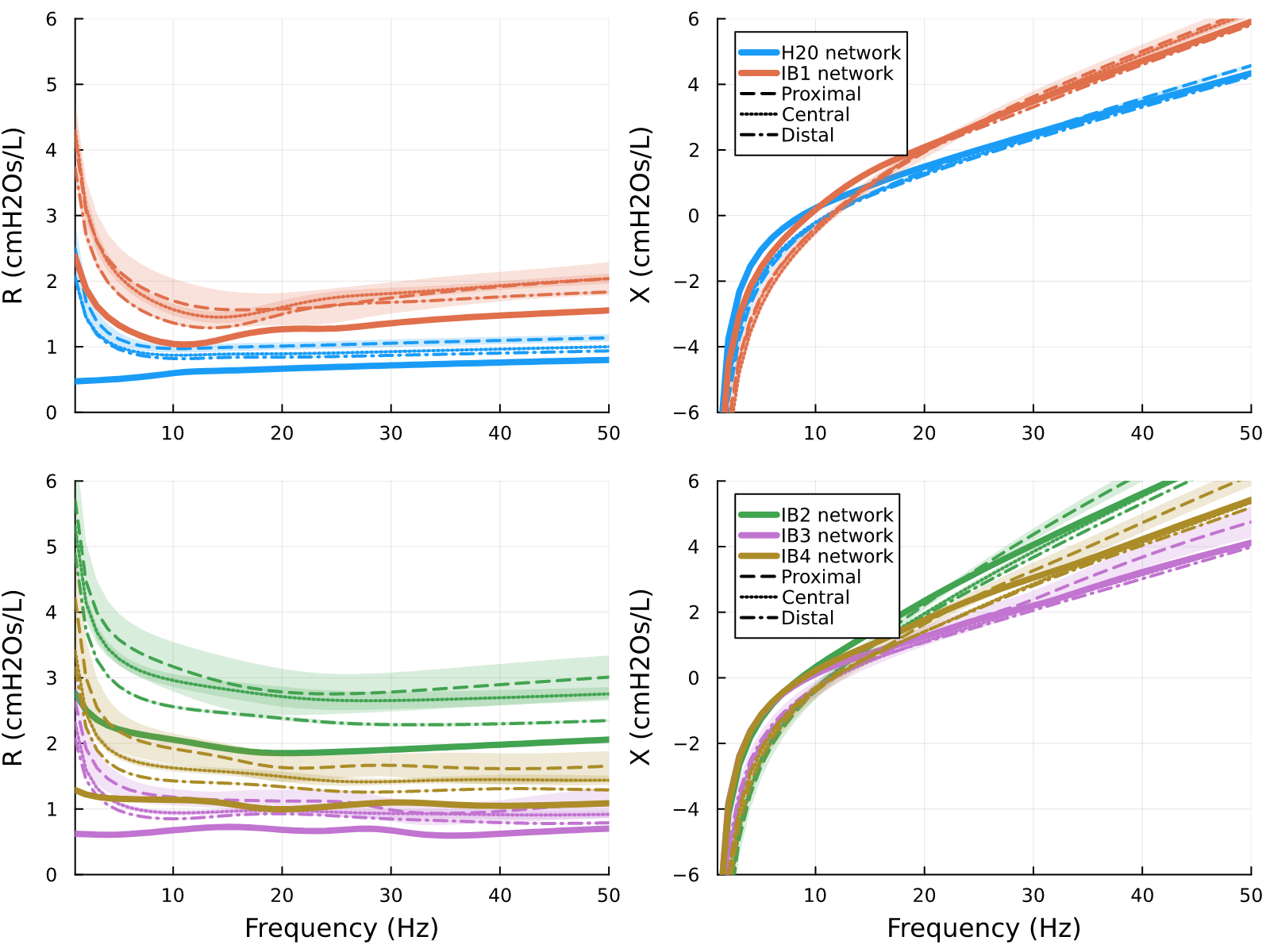}
\caption{Left: Plots of the resistance $R$ versus frequency for the 5 different baseline networks  (solid lines), consisting of 4 based on CT images (IB1--4) and 1 based on the Horsfield model \cite{horsfield_models_1971}. The dashed lines show the same networks with 50\% of either the proximal, central or distal airways constricted in that network (as defined in section \ref{sec:real_networks}). Right: Reactance $X$ versus frequency for the same networks. In the constricted cases, the curves are the mean of 20 realisations and the shaded regions show one standard deviation from the mean.}
\label{fig:impedance_depth}
\end{figure}

Figure \ref{fig:impedance_depth} shows the overall airway network resistance and reactance for frequencies between 1 and 50Hz. Ignoring the constricted cases for now, we can see that the image-based networks show some interesting fluctuations in resistance as a function of frequency that are not seen in the more idealised Horsfield network (labelled "H20") \cite{horsfield_models_1971}. This is most likely because there are regional differences in airway dimensions, more variability in airway length within a generation and greater branching asymmetry. Note that, as the constrictions are randomly placed and so regenerating the constrictions results in a different overall impedance each time, we calculate the impedance for 20 realisations of each constricted case. Note that because there is a very large number of distal airways, the 50\% selected at random are statistically representative of the whole collection and so there is little deviation of the predicted impedance in these cases. On the other hand, there is much more uncertainty in response when constricting the proximal airways, since they are fewer in number and more variable (as many are derived directly from the original CT image). 

These points aside, as would be expected, the constrictions increase the airway resistance significantly. Nonetheless, there is not a clear difference in frequency dependence between central, proximal or distal constrictions of either the reactance or resistance that would be an obvious signature of the location of airway narrowing. Airway narrowing itself appears to be somewhat captured by R5-R20, as suggested by other studies \cite{foy_characterising_2020}. However, variability between the networks means that the curves all have distinctly different shapes, suggesting that R5-R20 may not always capture the effect of constrictions, and that parameter inference may be a more reliable way to ascertain this information. Clearly, some improvements are required to this model, including better approximations of lung compliance and the impedance of other lung components before any generalisations could be made from these simulations. 

\subsection{Parameter identifiability in a simplified lung model}
\label{sec:param_ident}

\begin{figure}[h!]
\centering
\includegraphics[width=0.8\textwidth]{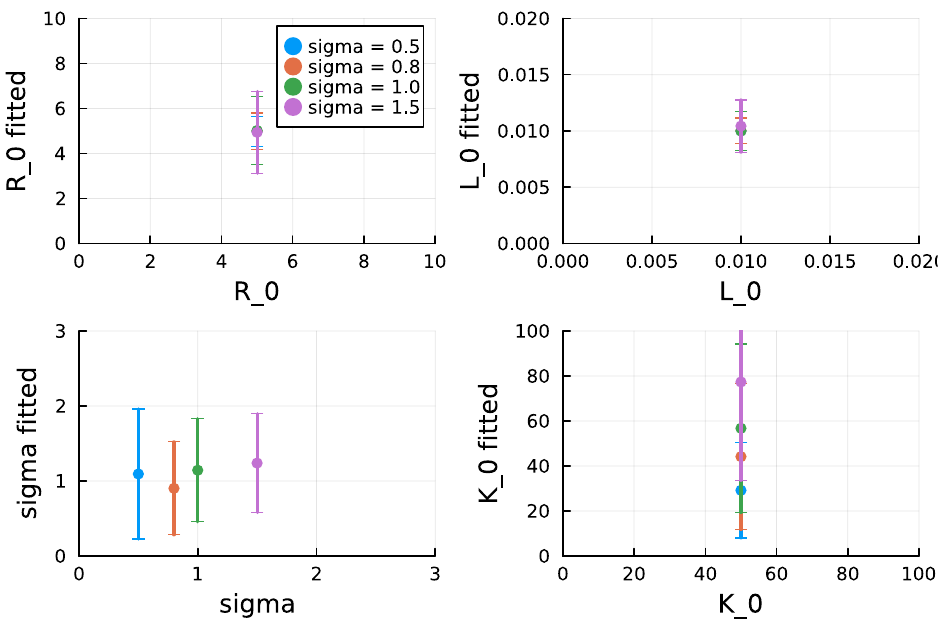}
\caption{Results of 4 parameter identification tests. In all tests the reference values were $\{K_0,R_0,L_0\} = \{50,5,0.01\}$ for the homogeneous parameters and the heterogeneity parameter was $\sigma \in \{0.5,0.8,1.0,1.5\}$. Each dot shows the posterior median value predicted, while the error bars show the 95\% confidence intervals.}
\label{fig:param_ident}
\end{figure}

We tested parameter identifiability by generating $Z$ for a small number of model runs and then fitting that data back to the same model. Figure \ref{fig:param_ident} shows that we get good convergence agreement for $R_0$ and $L_0$ parameters, but that $K_0$ and $\sigma$ are not identifiable. 

\begin{figure}[h!]
\centering
\includegraphics[width=0.5\textwidth]{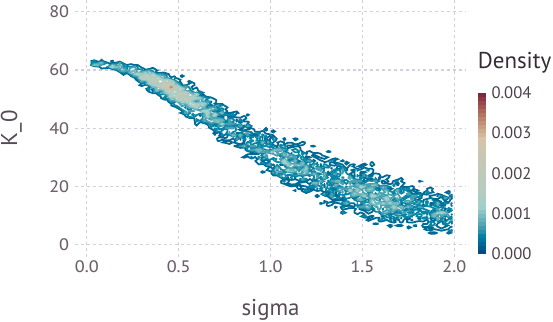}
\caption{Joint posterior distribution for the parameters $K_0$ and $\sigma$ in a parameter identification simulation with reference values $\sigma=0.5$ and $K_0 = 50$.}
\label{fig:K_0_sig_post}
\end{figure}

The joint posterior of these parameters shows a strong negative correlation because increasing(decreasing) $K_0$ has an indistinguishable effect as proportionately increasing(decreasing) $\sigma$ (figure \ref{fig:K_0_sig_post}). This leads us to the conclusion that, given only data on lung impedance, this simplified model would only be able to predict heterogeneity in the characteristic timescales for each unit (i.e. $(c_i K_0)^{-1}$) and not separate the elastic and resistance heterogeneity effects.

Note that we see that the posterior means of $K_0$ do change whereas the posterior means of $\sigma$ are $\approx 1$ in all cases. This is essentially because the priors on sigma are more restrictive, in effect, than the priors on $K_0$. Effectively, it appears only the product $\sigma K_0$ can be inferred and so the posterior distribution of $\sigma$ is almost unchanged from its prior, whereas $K_0$ is more constrained.

\subsection{Spectral properties of airway network impedance}
\label{sec:eigenmodes}

\begin{figure}
    \centering
    \includegraphics[width = 0.75\textwidth]{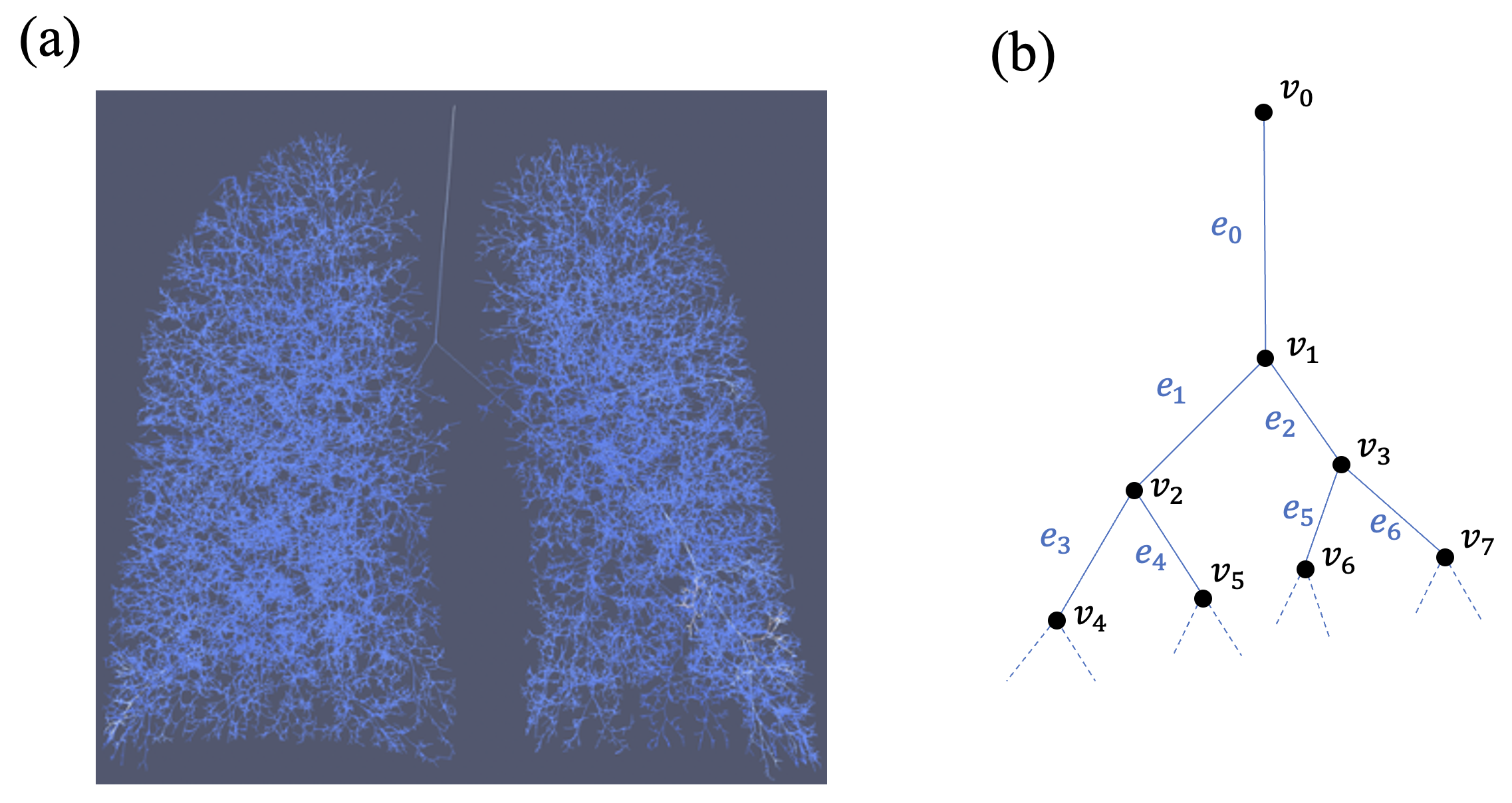}
    \caption{Airway network geometry that could be used for the normal mode calculation (Appendix \ref{sec:network-models}). (a) All airways in a whole-lung geometry, each represented by a single straight edge. (b) Schematic of part of a network, showing that the edges, $\{e_j\}$, and the vertices, $\{v_k\}$, are labelled. Physical properties such as airway wall stiffness, airway resistance, as well as air flow rates and pressures are defined on either the set of edges or on the set of vertices. }
    \label{fig:normal-modes}
\end{figure}

In Appendix \ref{sec:network-models}, we propose a method for computing the normal modes of a heterogeneous airway network. These normal modes provide the set of frequencies at which an imposed oscillation excites a response at the same frequency in the network. To calculate them, the lung geometry first has to be defined: it consists of a network of one-dimensional edges (see figure \ref{fig:normal-modes}), each representing a single airway, connected by a set of vertices, which either represent an airway bifurcation or, if the vertex is at a terminal end of the network, it represents an acinus. A vector of airway resistances, $\{R_j\}$, is defined on the edges of the network, and a vector of stiffnesses, $\{E_k\}$, is defined on the vertices, representing either the stiffness of the surrounding tissue at airway bifurcations or the (generally, much lower) stiffnesses of the acini. After then writing down equations relating the pressures at each vertex, the flow through each edge and the physical properties described above, an eigenvalue problem can be defined (see Appendix \ref{sec:network-models}), the solutions to which are the normal modes of the system. The normal-mode eigenvalues, $\{\omega_n^2\}$, are the squares of the oscillation frequencies, and the corresponding eigenmodes, $\{\mathsf{q}_n\}$, which are vectors on the vertices of the network, provide information about how different regions of the lungs are contributing to the total response when excited at each of the normal-mode frequencies. This information could be exploited in oscillometry to understand in more detail the spatial patterns of response in the lungs, and how these are affected by changes in airway resistance or compliance in disease. Appendix \ref{sec:network-models} also outlines how airway wall inertia or viscous damping in the wall could be incorporated into the analysis, and what the expected response of the system would be to forcing at a frequency close to, but not exactly equal to, one of the normal-mode frequencies, $\omega_n$.

One of the challenges in implementing this approach is defining an accurate airway network geometry, including the resistances and stiffnesses. Defining patient-specific geometries is potentially difficult, but there are pre-existing geometries in the literature (e.g., see \cite{whitfield2020}) derived from CT scans combined with space-filling algorithms for the small airways \cite{bordas_2015_developmet}, from which airway resistances can be approximated via Poiseuille's law, and there are existing empirical formulae that could be used to approximate airway wall stiffnesses \cite{kim2015}. Even if having patient-specific geometries is unlikely, having a smaller number of model geometries could be useful for understanding typical responses to oscillometry during the development process of a device. 


\section{Discussion} 
\label{sec:disc}

This report touches on a number of different approaches and ideas around modelling of lung oscillometry data. Sections \ref{sec:device_data} and \ref{sec:device_data_results} more directly address the challenge of calibrating models to account for the device parameters. Combining knowledge of the device with algorithms for parameter inference and data from the device alone, we were able to calibrate a minimal circuit model of the device to device-only data. The results were sensible, although the algorithms used were potentially too computationally expensive for the task at hand. Fitting the entire time-series data meant that no data was lost, however these results should be compared against fitting the response at the forced frequencies only to see if this changes the fit at all. If not then the latter would be more computationally efficient. This calibrated model could be used in tandem with models of the lung, chest and extra-thoracic components to interpret the data, and account for the contribution of the device and physical interactions between the device and the other components. 

In section \ref{sec:impedance_real} we calculated the effect that random airway blockages have on the overall impedance of realistic airway networks. This showed that, while the blockages had an clear effect on the resistance at low frequencies (as shown by previous studies) the effect of blocking at different depths were not noticeably distinct. Furthermore, different networks responded to different extents of blockage differently, so quantifying the extent of blockage from oscillometry data alone is unlikely to be possible. 

In section \ref{sec:param_ident} we trialled a simple inverse model of resistance heterogeneity in the lung to test whether this was an identifiable parameter from oscillometry data. Even with a minimal lung model, excluding extra-thoracic and device components, we found that the combination of resistance heterogeneity and compliance had an identifiable effect, but neither one was individually identifiable. 

Finally, section \ref{sec:eigenmodes} and appendix \ref{sec:network-models} looked in more detail at the phenomena of resonant frequencies of the lung airways and their impact on oscillometry measurements. 

This report outlines a number of interesting potential research directions inspired by the challenge of robustly analysing and interpreting oscillometry data. However, there are also limits to the model-based approaches outlined here. Using inappropriate or over-simplified models of the respiratory system can mean that the inferred parameters may not be representative of the physiological feature they are meant to quantify. Model selection algorithms are a potential approach to choosing the best model of several options, or alternatively parameter-free inference methods may be more appropriate for some aspects of the system. Models with too many features or parameters are subject to overfitting (or poor parameter identifiability) when there is not enough information in the data to specify all of the parameters. However, there is an exciting opportunity, particular for multi-modal devices such as the Respicorder, to combine data from multiple measurements modalities into a single model that is more representative of overall lung (patho)physiology than any individual measurement.

\section*{Author Contributions}
\addcontentsline{toc}{section}{Author Contributions}

All authors participated in the study group and contributed to the ideas and results discussed in this report. GRD provided the original study group challenge. OEJ, SM, SKN, MJR, SS, JS, LW and CAW all contributed to the drafting of this report. BSB, GD, and CAW contributed the editing and production of the final draft.

\label{EndOfText}
\newpage
\pagenumbering{Roman} 
\addcontentsline{toc}{section}{List of Figures}
\fancyfoot[C]{Page \thepage\ of \pageref{endOfDoc}}
\listoffigures
\thispagestyle{fancy}



\newpage
\addcontentsline{toc}{section}{References}
\bibliography{document.bib} 
\bibliographystyle{ieeetr}

\newpage
\section*{Appendices}
\addcontentsline{toc}{section}{Appendices}

\appendix

\section{Accessing normal modes of the lung} \label{sec:network-models}

The following theory provides an approach for computing the normal modes of a heterogeneous airway network, which may be exploited by forced oscillometry.

We represent the airway network as a directed graph with $N_v$ vertices (labelled by $k$) and $N_e$ edges (labelled by $j$).  The network topology is represented using the signed incidence matrix $A_{jk}$, such that $A_{jk}=1$ where edge $j$ points into vertex $k$, $A_{jk}=-1$ where edge $j$ points out of vertex $k$ and $A_{jk}=0$ otherwise.  The network has a single inlet at vertex zero, which supplies edge 0.  Outlet vertices represent acini.  Internal vertices can be associated with airway bifurcations.  Each edge is assigned a fixed length $L_j$ and cross-sectional area $a_j$ (which we assume is also constant).  Each vertex is assigned a volume $V_k(t)$, so that the total lung volume is $\sum_k V_k(t)$, where $t$ is time. We define pressures $p_k(t)$ on vertices and fluxes $q_j(t)$ along edges.  

Airways are assigned viscoelastic material properties at vertices, such that 
\begin{equation}
    p_{k,t}=E_k V_{k,t}+D_k V_{k,tt} + M_k V_{k,ttt}.
    \label{eq:con}
\end{equation}
We expect the airway stiffness $E_k$ to be large at bifurcations connecting conducting airways, but relatively small in peripheral units (to allow alveolar expansion and contraction).  $D_k$ represents viscous damping in airway walls and $M_k$ represents airway wall inertia.  

Mass conservation for incompressible air flow is expressed for vertex $k$ as 
\begin{equation}
    V_{k,t}=-{\textstyle \sum_j} A_{jk}q_j+S_k.
    \label{eq:mas}
\end{equation}
The $S_k$ allow boundary conditions to be applied on the network.  $S_0$ defines a source/sink at the inlet.  We set $S_k=0$ at all internal vertices.  Setting $S_k=0$ at outlet vertices forces acinar volumes to change as flow enters and leaves them via (\ref{eq:con}).

The momentum equation describing unidirectional flow along the airway associated with edge $j$ is
\begin{equation}
    \frac{\rho}{a_j}q_{j,t}+{\textstyle{\sum_k}} A_{jk}\frac{p_k}{L_j}=-R_j q_j.
    \label{eq:mom}
\end{equation}
Here, $\rho$ is the air density, assumed constant, and $R_j$ models viscous resistance (assuming a simple linear dependence on flow rate).  The second term in (\ref{eq:mom}) is the pressure gradient along the airway.

Differentiating (\ref{eq:mom}) with respect to time and using (\ref{eq:con}, \ref{eq:mas}) gives
\begin{equation}
\frac{\rho}{a_j}q_{j,tt}+{\textstyle{\sum_{k,j'}}} \frac{A_{jk}}{L_j}(E_k+D_k \partial_t+M_k \partial_t^2)(S_k-A_{j'k} q_{j'})=-R_j q_{j,t}.
\end{equation}
For the time being we set $D_k=0$ and $M_k=0$.  Then we seek $q_j(t)$ satisfying
\begin{equation}
\frac{\rho}{a_j}q_{j,tt}+R_j q_{j,t}+{\textstyle{\sum_{k,j'}}} \frac{1}{L_j}A_{jk}E_k(S_k-A_{j'k} q_{j'})=0.
\label{eq:wa1}
\end{equation}
Defining the matrices $\mathsf{E}=\mathrm{diag}(E_1, \dots, E_{N_v})$, $\mathsf{R}=\mathrm{diag}(R_1\dots,R_{N_e})$, $\mathsf{L}=\mathrm{diag}(L_1\dots,L_{{N_e}})$, $\mathsf{a}=\mathrm{diag}(a_1,\dots,a_{{N_e}})$, $\{\mathsf{A}\}_{jk}=A_{jk}$, and the vectors $\mathsf{q}=(q_1,\dots,q_{N_e})^\top$, $\mathsf{S}=(S_1,\dots, S_{N_v})^\top$, we can write (\ref{eq:wa1}) in matrix form as
\begin{equation}
{\rho}\mathsf{a}^{-1} \mathsf{q}_{tt}+\mathsf{R} \mathsf{q}_{t}+ \mathsf{L}^{-1} \mathsf{A} \mathsf{E} (\mathsf{S}-\mathsf{A}^\top \mathsf{q})=\mathsf{0}.
\label{eq:wa2}
\end{equation}

We then write $S_k=\hat{S}_k e^{i\omega t}$, $q_j=\hat{q}_j e^{i\omega t}$, $p_k=\hat{p}_k e^{i\omega t}$, assuming the forcing at frequency $\omega$ drives a response with the same frequency, so that (\ref{eq:wa2}) becomes
\begin{equation}
-{\rho}\omega^2 \mathsf{a}^{-1} \hat{\mathsf{q}}+i\omega \mathsf{R} \hat{\mathsf{q}}+ \mathsf{L}^{-1} \mathsf{A} \mathsf{E} (\hat{\mathsf{S}}-\mathsf{A}^\top \hat{\mathsf{q}})=\mathsf{0}.
\label{eq:wa3}
\end{equation}
This provides a linear problem for $\hat{\mathsf{q}}$ driven by the forcing $\hat{\mathsf{S}}$.  (In a Fourier decomposition such as this, some authors introduce additional frequency dependence to $\mathsf{R}$.)

The corresponding pressure is determined by differentiating (\ref{eq:con}) and using (\ref{eq:mas}, \ref{eq:mom}), to give 
\begin{equation}
    p_{k,tt}=E_k\left[S_{k,t}+{\textstyle \sum_j}A_{jk}\left(\frac{a_j R_j}{\rho}q_j +\frac{a_j}{\rho}{\textstyle{\sum_{k'}}} A_{jk'} \frac{p_{k'}}{L_j} \right)\right].
\end{equation}
In vector form, this becomes
\begin{equation}
-\rho \omega^2 \mathsf{I} \hat{\mathsf{p}} =\rho i \omega \mathsf{E} \hat{\mathsf{S}}+ \mathsf{E} \mathsf{A}^\top\mathsf{a}  \left( \mathsf{R} \hat{\mathsf{q}} +\mathsf{L}^{-1} \mathsf{A} \hat{\mathsf{p}}  \right).
\label{eq:wa4}
\end{equation}
Steady ventilation (setting $\omega=0$) is described from (\ref{eq:wa2}, \ref{eq:wa4}) by
\begin{equation}
    \hat{\mathsf{S}}=\mathsf{A}^\top\hat{\mathsf{q}},\quad \mathsf{R}\hat{\mathsf{q}}+\mathsf{L}^{-1}\mathsf{A}\hat{\mathsf{p}}=\mathsf{0},
\end{equation}
giving a Poisson problem for the pressure
\begin{equation}
    -\mathsf{A} (\mathsf{L}\mathsf{R})^{-1}\mathsf{A}^\top \hat{\mathsf{p}}=\hat{\mathsf{S}}
    \label{eq:vent}
\end{equation}
that involves the Laplacian operator investigated in \cite{whitfield2020}.

When the flow is unsteady, we can investigate (\ref{eq:wa3}) in isolation.  It is instructive to consider the case in which airway resistance is weak.  In the absence of forcing, the flow rate satisfies the eigenvalue problem
\begin{equation}
-  \mathsf{a} \mathsf{L}^{-1} \mathsf{A} \mathsf{E} \mathsf{A}^\top \hat{\mathsf{q}}={\rho}\omega^2 \hat{\mathsf{q}},
\label{eq:ev}
\end{equation}
which involves a different type of Laplacian operator to that appearing in (\ref{eq:vent}).  The eigenfunctions $\hat{\mathsf{q}}_n$ and eigenvalues $\rho \omega_n^2$ of (\ref{eq:ev}) reveal the normal modes of the lung.  We define the inner product $\langle \mathsf{g} , \mathsf{h} \rangle=\mathsf{g}^\top \mathsf{L}\mathsf{a}^{-1} \mathsf{h}$ for any scalar fields $\mathsf{g}$, $\mathsf{h}$ defined on edges.  Then 
\begin{equation}
    \langle \hat{\mathsf{q}}_n,-\mathsf{a}\mathsf{L}^{-1}\mathsf{A} \mathsf{E}\mathsf{A}^\top \hat{\mathsf{q}}_m\rangle = 
- \hat{\mathsf{q}}_n^\top \mathsf{A} \mathsf{E}\mathsf{A}^\top \hat{\mathsf{q}}_m
=- \hat{\mathsf{q}}_m^\top \mathsf{A} \mathsf{E}\mathsf{A}^\top \hat{\mathsf{q}}_n
  =  \langle \hat{\mathsf{q}}_m,-\mathsf{a}\mathsf{L}^{-1}\mathsf{A} \mathsf{E}\mathsf{A}^\top \hat{\mathsf{q}}_n\rangle,
\end{equation}
implying that 
\begin{equation}
    (\omega_n^2-\omega_m^2)\langle \hat{\mathsf{q}}_n,\hat{\mathsf{q}}_m\rangle=0,
\end{equation}
demonstrating orthogonality of the eigemodes.  It would be of interest to compute the eigenmodes for a heterogeneous airway network, in order to assess the spatial structure of the individual modes.  One can expect modes of different frequencies to excite different regions of the airway network: certain modes will access smaller airways better than others.

We can then imagine imposing a forcing that is close to a frequency $\omega_n$.  Write 
\begin{equation}
    \omega=\omega_n+\tilde{\omega},\quad \hat{\mathsf{q}}=B\hat{\mathsf{q}}_n+\tilde{\mathsf{q}},
    \label{eq:exp}
\end{equation}
where $\tilde{\omega}$ is the detuning, $B$ is a complex amplitude (to be determined) and tildes denote perturbations.  We substitute (\ref{eq:exp}) into (\ref{eq:wa2}) and linearise, retaining resistance and forcing terms, to obtain
\begin{equation}
    (-\rho\omega_n^2 -\mathsf{a}\mathsf{L}^{-1}\mathsf{A}\mathsf{E}\mathsf{A}^\top)\tilde{\mathsf{q}} =2\rho \omega_n \tilde{\omega} B\hat{\mathsf{q}}_n-i\omega_n \mathsf{a} \mathsf{R}B\hat{\mathsf{q}}_n-\mathsf{a}\mathsf{L}^{-1} \mathsf{A}\mathsf{E} \hat{\mathsf{S}}.
\end{equation}
The forcing on the right must be orthogonal to the kernel of the operator on the left under the inner product, requiring 
\begin{equation}
0=\langle \hat{\mathsf{q}}_{n}, 2\rho \omega_n \tilde{\omega} B\hat{\mathsf{q}}_n-i\omega_n \mathsf{a} \mathsf{R}B\hat{\mathsf{q}}_n-\mathsf{a}\mathsf{L}^{-1} \mathsf{A}\mathsf{E} \hat{\mathsf{S}}\rangle
\end{equation}
which implies
\begin{equation}
    B=\frac{\langle \hat{\mathsf{q}}_n,\mathsf{a}\mathsf{L}^{-1} \mathsf{A}\mathsf{E} \hat{\mathsf{S}}\rangle}{2\rho \omega_n \tilde{\omega} \langle\hat{\mathsf{q}}_n,\hat{\mathsf{q}}_n\rangle -i\omega_n     \langle\hat{\mathsf{q}}_n,\mathsf{a}\mathsf{R} \hat{\mathsf{q}}_n\rangle}.
\end{equation}
The magnitude of the response to forcing close to a resonant frequency is therefore given by 
\begin{equation}
    B=\frac{\langle \hat{\mathsf{q}}_n,\mathsf{a}\mathsf{L}^{-1} \mathsf{A}\mathsf{E} \hat{\mathsf{S}}\rangle}{\sqrt{(2\rho \omega_n \tilde{\omega} \langle\hat{\mathsf{q}}_n,\hat{\mathsf{q}}_n\rangle)^2 +(\omega_n     \langle\hat{\mathsf{q}}_n,\mathsf{a}\mathsf{R} \hat{\mathsf{q}}_n\rangle)^2}}.
\end{equation}
This demonstrates how the maximum response (when $\tilde{\omega}=0$) is regulated by viscous resistance.  The numerator shows that a large response also requires the relevant mode to have an appreciable amplitude near the source.

Measuring the amplitude and phase of the response to forcing across a range of frequencies can be expected to yield resonant responses near certain frequencies.  Using these responses to determine properties of the lung requires \textit{a priori} knowledge of the modal structure.  However this model could be used to assess the impact of localised airway blockages on the response to forcing.


\label{endOfDoc}
\end{document}